\documentclass[3p,sort&compress]{elsarticle}

\usepackage{epsfig} 
\usepackage{lineno,hyperref}
\modulolinenumbers[5]
\usepackage{amssymb}
\usepackage{amsmath}
\usepackage{caption}

\journal{Physics Letters B}

\begin{document}

\begin{frontmatter}

\title{Halo structure of $^8$B determined from intermediate energy proton elastic scattering in inverse kinematics}

\author[address1]{G.A.~Korolev\corref{mycorrespondingauthor}}
\cortext[mycorrespondingauthor]{Corresponding author}
\ead{Korolev\_GA@pnpi.nrcki.ru}

\author[address1]{A.V.~Dobrovolsky} 
\author[address1]{A.G.~Inglessi}
\author[address1]{G.D.~Alkhazov} 
\author[address2]{P.~Egelhof}
\author[address2]{A.~Estrad\'{e}}
\author[address2]{I. Dillmann}
\author[address2]{F. Farinon}
\author[address2]{H. Geissel}
\author[address2]{S. Ilieva}
\author[address2]{Y.~Ke}
\author[address1]{A.V. Khanzadeev} 
\author[address2]{O.A. Kiselev}
\author[address2]{J. Kurcewicz}
\author[address2]{X.C. Le}
\author[address2]{Yu.A.~Litvinov}
\author[address1]{G.E. Petrov}
\author[address2]{A. Prochazka}
\author[address2]{C. Scheidenberger}
\author[address1]{L.O.~Sergeev}
\author[address2]{H. Simon}
\author[address2]{M. Takechi}
\author[address2]{S. Tang}
\author[address2]{V. Volkov}
\author[address1]{A.A.~Vorobyov}
\author[address2]{H.~Weick}
\author[address1]{V.I. Yatsoura}

\address[address1]{Petersburg Nuclear Physics Institute, National Research Centre Kurchatov Institute, Gatchina, 188300 Russia}
\address[address2]{GSI Helmholtzzentrum f\"{u}r Schwerionenforschung GmbH 64291 Darmstadt, Germany}

\begin{abstract}
The absolute differential cross section for small-angle proton elastic scattering on the proton-rich $^8$B nucleus has been measured in inverse kinematics for the first time. The experiment was performed using a secondary radioactive beam with an energy of 0.7~GeV/u at GSI, Darmstadt. The active target, namely hydrogen-filled time projection ionization  chamber IKAR, was used to measure the energy, angle and vertex point of the recoil protons. The scattering angle of the projectiles was simultaneously determined by the tracking detectors. The measured differential cross section is analyzed on the basis of the Glauber multiple scattering theory using phenomenological nuclear-density distributions with two free parameters. The radial density distribution deduced for $^8$B exhibits a halo structure with the root-mean-square (rms) matter radius $R_{\rm m} = 2.58 (6)$~fm and the rms halo radius $R_{\rm h} = 4.24 (25)$~fm. The results on $^8$B are compared to those on the mirror nucleus $^8$Li investigated earlier by the same method. A comparison is also made with previous experimental results and theoretical predictions for both nuclei.
\end{abstract}


\end{frontmatter}


The study of the nuclear structure of unstable exotic nuclei has become an important direction of investigations in nowadays nuclear physics~\cite{Tanihata2013,Jonson2004}.
A characteristic feature of some light weakly bound nuclei is that they may form a neutron or a proton halo with a dilute mass distribution extending far outside of a compact core of the nucleus. The proton drip-line nucleus $^8$B has received much attention from both theoretical and experimental points of view. With a proton separation energy of 0.138~MeV, $^8$B is the most likely candidate for having a proton halo structure. In addition, $^8$B plays an essential role in the solar neutrino problem. The $^8$B nucleus is produced in the sun through the $^7$Be($p,\gamma$)$^8$B reaction and emits a high energy neutrino~\cite{Bahcal1982}. The proton capture rate in $^7$Be strongly depends on the $^8$B structure. Thus the size of $^8$B and the shape of the proton density distribution at large distances are important for the description of the solar neutrino flux~\cite{Bahcal1982,Riisager1993}.

At present, $^8$B is considered to be a proton halo nucleus, in spite of the existence of the Coulomb and centrifugal barriers. 
Experimentally, the halo structure of $^8$B was suggested by the Osaka group~\cite{Minamisono1992} to explain 
the unusually large quadrupole moment of this nucleus as compared to the value for the mirror nucleus $^8$Li. 
However, theoretical calculations~\cite{Nakada1994} have shown that the large quadrupole moment of $^8$B 
can be explained without the existence of a proton halo. The main evidence for the proton halo structure in $^8$B came 
from the experiments in which the narrow longitudinal momentum distribution of $^7$Be fragments after proton break-up 
and the large one-proton removal cross-sections in break-up reactions were measured~\cite{Schwab1995,Kelley1996,Smedberg1999}. 
The size and the shape of the radial distribution of the nuclear matter are fundamental properties of nuclei and can be 
the most convincing evidence for the proton halo structure. 
The root-mean-square (rms) matter radius $R_{\rm m}$ of $^8$B was deduced in several experiments through measurements 
of the reaction (interaction) cross section $\sigma_{\rm R}$ ($\sigma_{\rm I}$)~\cite{Tanihata88,Khali96,Obuti96,Negoita96,Fukuda99,Fan15}. 
However, the values obtained for the matter radius from these experiments  are widely scattered, ranging from $2.38 (2)$~fm to $2.61 (8)$~fm.

The proton-nucleus elastic scattering at intermediate energies is considered to be one of the best methods to obtain nuclear matter density distributions in stable nuclei~\cite{Tanihata2013,Alk78}. At these energies, the Glauber multiple scattering theory accurately describes the process of elastic scattering and connects the measured differential cross section with the nuclear matter distribution in a rather unambiguous way~\cite{Alk78}. In order to study exotic nuclei, it was proposed and later realized~\cite{Alk92,Alk97} to perform experiments in inverse kinematics using radioactive nuclear beams and the hydrogen active target IKAR. As theoretical considerations have shown~\cite{Alk92}, proton scattering at small momentum transfers is particularly sensitive to the nuclear matter radius and to the halo structure of nuclei. Indeed, scattering on halo nucleons contributes to the slope of the differential cross sections d$\sigma$/d$t$ at low momentum transfers $|t|$, that means at small scattering angles. An analysis of the shape of the measured cross sections makes it possible to determine the sizes of the nuclear core and of the halo. The proposed method was successfully used at GSI Darmstadt at energies around 700~MeV/u to measure absolute differential cross sections for proton elastic scattering in inverse kinematics on the radioactive neutron-rich isotopes $^6$He, $^8$He, $^8$Li, $^9$Li,$^{11}$Li, $^{12}$Be and $^{14}$Be~\cite{Neumaier02,Alk02,Egelhof02,Dobrov06,Ilieva12}. An analysis of the data yielded parameters of the nuclear matter distributions. The elastic $p^4$He and $p^6$Li differential cross sections were also measured as a consistency check of the experimental method, including the procedure applied for the data analysis~\cite{Alk02,Dobrov06}.

In this Letter, we present the first measurement of the absolute differential cross section for proton elastic scattering on the proton-rich $^8$B nucleus in inverse kinematics at an energy of 0.7~GeV/u.

The experiment was carried out at the radioactive-beam facility of GSI, Darmstadt. A~ primary $^{22}$Ne beam delivered from the heavy-ion synchrotron (SIS) was focused on an 8~g/cm$^2$ Be production target at the entrance of the FRagment Separator (FRS). The produced boron ions were separated according to their magnetic rigidity, and due to their nuclear charge by inserting an achromatic (2.7~g/cm$^2$) aluminum degrader at the dispersive central focal plane. The contamination from other nuclei was below the 0.1$\%$ level. The energy of the secondary beam at the centre of the hydrogen target was 699~MeV/u with an energy spread of 1.3$\%$. The mean energy value was determined with an accuracy of about 0.1$\%$. The beam intensity was $\sim3\cdot10^3$ s$^{-1}$.

\begin{figure} [htbp]
	\centering
	\psfig{file=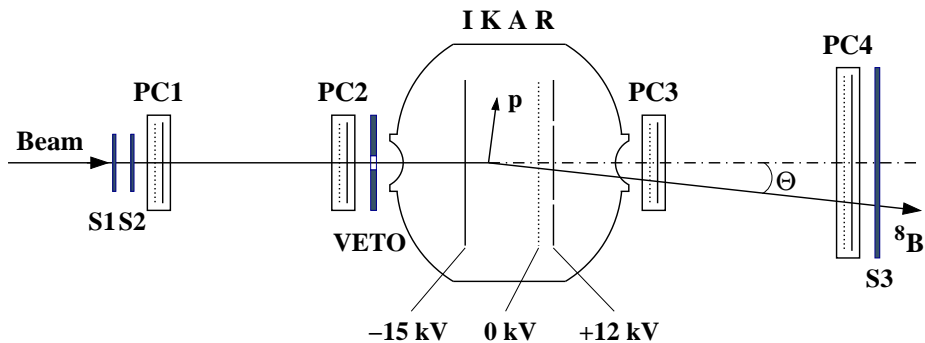,scale=0.80} 
	\caption{Schematic view of the experimental set-up for small-angle proton elastic scattering on exotic nuclei in inverse kinematics. The hydrogen-filled ionization chamber IKAR serves simultaneously as a gas target and a detector for recoil protons (for details see Ref.~\cite{Neumaier02}). For the sake of simplicity only one chamber module of six identical ones is shown. The tracking system consisting of four multi-wire proportional chambers PC1--PC4 determines the scattering angle $\theta$ of the projectile. The scintillation counters S1--S3 and VETO are used for beam identification and triggering.} 
	\label{setup}
\end{figure}

A schematic view of the experimental layout is shown in Fig.~\ref{setup}. The main constituent of the set-up was the active target IKAR filled with pure hydrogen at a pressure of 10 bar, which served simultaneously as a gas target and a recoil proton detector. IKAR was developed at PNPI~\cite{Vor74,Vor82,Burq78} and was originally used in experiments on small-angle hadron elastic scattering. The chamber consists of six identical modules. The signals from the ionization chamber provide the recoil energy $T_{\rm R}$, the recoil angle and the vertex point $Z_{\rm V}$ of the interaction. The recoil protons were registered in IKAR in coincidence with the scattered $^8$B particles. The momentum transfer could be determined either from the measured recoil energy $T_{\rm R}$ or from the value of the scattering angle $\theta$ of the projectiles which was measured by a tracking detector system consisting of 2 pairs of two-dimensional multi-wire proportional chambers (PC1--PC2 and PC3--PC4), arranged upstream and downstream  with respect to IKAR. A set of scintillation counters (S1, S2 and S3) was used for triggering and identification of the beam particles via time-of-flight and d$E/$d$x$ measurements, while a circular-aperture scintillator VETO selected the projectiles which entered IKAR within an area with a diameter of 2~cm around the central axis. A high detection efficiency for beam particles and elastic-scattering events in IKAR insured the high accuracy of the absolute normalization of the measured cross section ($\sim2\%$). A detailed description of the experimental set-up is given in Ref.~\cite{Neumaier02}.

\begin{figure} [tp]
	\centering 
	\psfig{file=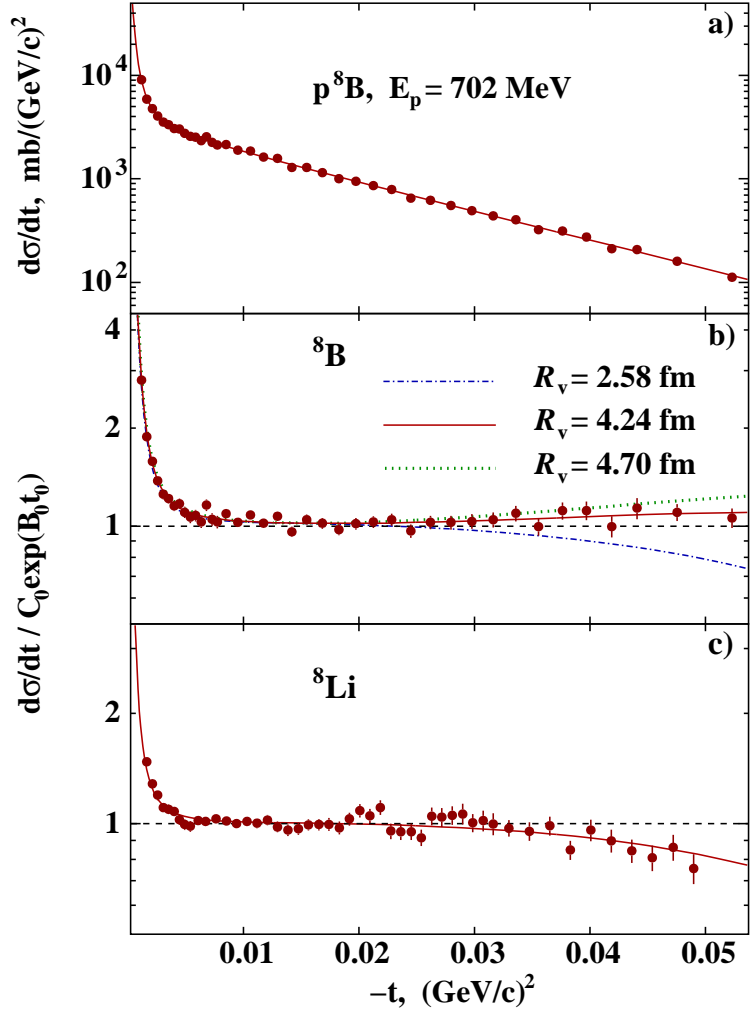,scale=0.65} 
	\caption{
		a) Absolute differential cross section for $p^8$B elastic scattering. The indicated energy corresponds to the equivalent proton energy for direct kinematics. The plotted error bars denote in all Figures a-c statistical errors only (in Fig.~\ref{crs}a the error bars are smaller than the symbols). The solid line represents the cross section calculated using the Glauber multiple-scattering theory using a phenomenological matter distribution GO with fitted parameters. The resulting matter radius of $^8$B is  $R_{\rm m} = 2.58$~fm. \\
		b) The same cross section divided by an exponential function (for details see text). The three presented curves are obtained using the GO parameterization with the same $R_{\rm m} = 2.58$~fm, but with different values of the valence proton radius $R_{\rm v}$. The case of $R_{\rm v}$ = 2.58~fm ($R_{\rm c}$ = 2.58~fm) corresponds to the absence of a halo and exhibits a negative curvature, the version with $R_{\rm v}$ = 4.24~fm ($R_{\rm c}$ = 2.24~fm) is the best fit to the experimental points, and the case of $R_{\rm v}$ = 4.70~fm ($R_{\rm c}$ = 2.11~fm) illustrates the sensitivity of the cross section to the change in the $R_{\rm v}$ value. The positive curvature indicates the halo structure of $^8$B. \\
		c) The cross section divided by an exponential function for $p^8$Li elastic scattering~\cite{Dobrov06}. The best fit curve with $R_{\rm m}$ = 2.50(6)~fm and $R_{\rm v}$ = 2.58(48)~fm ($R_{\rm c}$ = 2.48~fm) has a negative curvature and demonstrates a nonhalo structure for $^8$Li.} 
	\label{crs}
\end{figure}

The major steps in the data analysis, such as the selection of the elastic scattering events, were the same as in the previous experiments with the same method~\cite{Neumaier02,Alk02,Egelhof02,Dobrov06,Ilieva12}. The absolute differential cross section  d$\sigma/$d$t$ was determined using the relation 
\begin{equation}
{\rm d}\sigma / {\rm d}t = {\rm d}N /( {\rm d}t M n \Delta L)~.
\end{equation}
Here, ${\rm d}N$ is the number of elastic proton-nucleus scattering events in the interval ${\rm d}t$ of the four-momentum transfer squared, $M$ is the corresponding number of beam particles impinging on the target, $n$ is the density of the hydrogen nuclei known from the measured gas pressure and temperature, and $\Delta L$ is the effective target length. The value of $t$ was calculated as $|t| = 2 m T_{\rm R}$, (where $m$ is the mass of the proton) for the lower momentum transfers, and from the scattering angle $\theta$ of the projectiles for the higher momentum transfers~\cite{Ilieva12}. The differential cross section ${\rm d}\sigma/{\rm d}t$ obtained in this experiment in the $t$-range $0.001< |t|< 0.06$~(GeV/$c)^2$ is displayed in Fig.~\ref{crs}a. The indicated energy corresponds to the equivalent proton energy in direct kinematics. The uncertainty in the $t$-scale calibration was estimated to be about 1.5\%.

To establish the nuclear matter density distribution from the measured cross section, the Glauber multiple scattering theory was applied. Calculations were performed using the basic Glauber formalism for proton-nucleus elastic scattering and taking experimental data on the elementary proton-proton and proton-neutron scattering amplitudes as input (for details see Ref.~\cite{Alk02}). In the analysis of the experimental data, the nuclear many-body density $\rho_{\rm A}$ was taken as a product of the one-body densities, which were parameterized with different functions. The parameters of these densities were found by fitting the calculated cross section to the experimental data. The fitting procedure is described in detail in Ref.~\cite{Alk02}. In the analysis, four parameterizations of phenomenological nuclear density distributions were applied, labeled as SF (symmetrized Fermi), GH (Gaussian-halo), GG (Gaussian-Gaussian) and GO (Gaussian-oscillator). Each of these parameterizations has two free parameters. While the SF and GH parameterizations do not make any difference between core and halo distributions, the GG and GO parameterizations assume that the nuclei consist of core nucleons and valence nucleons with different spatial distributions. The core distribution is assumed to be a Gaussian one in both the GG and GO parameterizations. The valence nucleon density is described by a Gaussian or a $1p$ shell harmonic oscillator-type distribution within the GG or GO parameterization, respectively. The free parameters in the GG and GO parameterizations are the rms radii $R_{\rm c}$ and $R_{\rm v}$ ($R_{\rm h}$) of the core and valence (\textquotedblleft halo\textquotedblright) nucleon distributions. It was assumed that $^8$B consists of the $^7$Be core and a loosely bound valence proton (the core density was normalized to 7, while the halo density to 1).

In the data analysis, good descriptions of the cross sections have been obtained with all used density parameterizations. The value of $R_{\rm m}$, averaged over the results obtained with all density parameterizations is
\begin{equation}
{R_{\rm m} ~=~ 2.58 ~\pm 0.06 ~\rm {fm}}~,
\end{equation}
where the error includes statistical and systematical uncertainties. The systematical errors in $R_{\rm m}$ appear due to uncertainties in the absolute normalization of the cross sections, in the $t$-scale calibration and in the parameters of the elementary proton-nucleon scattering amplitudes (see Ref.~\cite{Alk02}). The systematic uncertainty in $R_{\rm m}$ arising due to different model density parameterizations used is $\sim0.026$ fm. The mean values for the core and halo radii of $^8$B deduced with both the GG and GO parameterizations are $R_{\rm c} = 2.25 (3)$~fm and $R_{\rm h} = 4.24 (25)$~fm, respectively. The relation between the rms radii can be written as
\begin{equation}
{A {R_{\rm m}}^2 = (A - 1) {R_{\rm c}}^2 + {R_{\rm h}}^2}~,
\end{equation}
where $A$ is the mass number. The solid line in Fig.~\ref{crs}a represents the result of the ${\rm d}\sigma/{\rm d}t$ calculations with the GO parameterization. At $|t| < 0.005$~(GeV/$c)^2$ the steep rise of the cross section with decreasing $|t|$ is caused by Coulomb scattering. The behavior of the measured curvature of the differential cross section at $0.005< |t| < 0.06$ (GeV/$c)^2$ is an indication of the halo occurrence. This effect can be seen if one plots the cross section divided by the exponential function $C_{\rm 0} \textrm{exp}(B_{\rm 0}t)$, where $B_{\rm 0}$ and $C_{\rm 0}$ are the slope and the absolute value of the nuclear part of the differential cross section calculated at $|t| = 0.01$ (GeV/$c)^2$. Such a plot is shown in Fig.~\ref{crs}b for the GO parameterization. The halo nuclei demonstrate a positive curvature in the $t$-dependence of $\textrm{ln} ({\rm d}\sigma/{\rm d}t)$~\cite{Alk02,Dobrov06}. This may be explained by the fact that contributions to the cross section for proton scattering from the core and from the halo of these nuclei have a different angular dependence. The contribution to the cross section from the scattering on the halo proton decreases faster with increasing $|t|$ than that from the scattering on the core nucleons. Note, that the cross section contribution from scattering on the nuclear halo is concentrated at low momentum transfers, whereas the scattering from the core contributes both at low and high momentum transfers. When we fit the calculated cross section to the experimental data, the fitted core size is dependent on the assumed halo size. Thus the halo size indirectly influences the behaviour of the calculated cross section at high momentum transfers. In Fig.~\ref{crs}b the sensitivity of the curvature in $\textrm{ln} ({\rm d}\sigma/{\rm d}t)$ to the structure of $^8$B is shown. The best fit to the experimental points corresponds to the curve with $R_{\rm v}$ = 4.24~fm and demonstrates the positive curvature, the negative curvature corresponds to the case of $R_{\rm v}$ = $R_{\rm m}$ = 2.58~fm. Also shown is the version with $R_{\rm v}$ = 4.70~fm.

In many theoretical investigations, the proton-rich $^8$B nucleus is described together with its mirror partner $^8$Li~\cite{Kitagawa93,Csoto93,Baye94,Varga95,Grigorenko98,Kitagawa99,Chandel03}. In particular, the nucleon structure of both these nuclei can be considered in a three-body approach~\cite{Csoto93,Baye94,Varga95,Grigorenko98}. The wave functions of $^8$Li and $^8$B display a high mirror symmetry. In Fig.~\ref{crs}c we show the cross section divided by an exponential function for elastic $p^8$Li scattering at the energy $E_{\rm p}$ = 698 MeV measured earlier with the same method~\cite{Dobrov06}. In the analysis, the $^8$Li nucleus was considered as consisting of the $^7$Li core and a valence neutron. The best-fit curve was also obtained with the GO parameterization, the values of the matter and valence neutron radii being very close to each other: $R_{\rm m} = 2.50 (6)$~fm and $R_{\rm v} = 2.58 (48)$~fm. Also the negative curvature of the fitted curve (see Fig.~\ref{crs}c) argues against a neutron halo structure in $^8$Li.

\begin{figure} [tp]
	\centering 
	\psfig{file=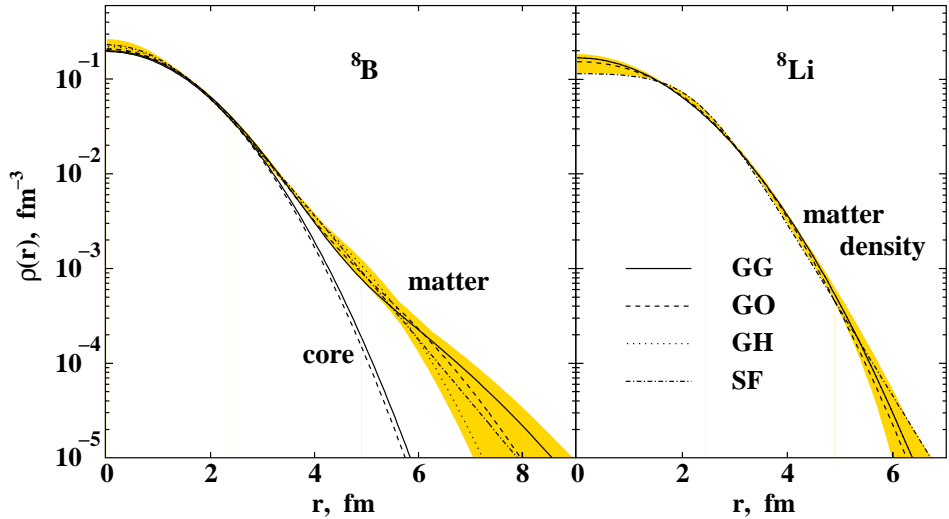,scale=0.50} 
	\caption{The core and nuclear point matter distributions deduced for $^8$B with the GG, GO, GH and SF parameterizations and the matter distribution of the mirror nucleus $^8$Li studied in the previous experiment~\cite{Dobrov06}. The shaded areas represent the envelopes of the density variation within the model parameterizations applied, superimposed by the statistical errors. All density distributions are normalized to the number of nucleons.} 
	\label{dens4}
\end{figure}

The core and matter distributions deduced for $^8$B by using different model parameterizations are compared in Fig.~\ref{dens4} with the matter distribution of $^8$Li~\cite{Dobrov06}. All density distributions refer to point-nucleon distributions. The results for all the four parameterizations for the description of the $^8$B matter density are rather similar and demonstrate a clear evidence for the proton halo existence. The experimental nuclear-matter radii of $^8$B and $^8$Li determined from the proton elastic scattering (this work and~\cite{Dobrov06}) and from reaction cross sections~\cite{Tanihata88,Khali96,Obuti96,Negoita96,Fukuda99,Fan15} are presented in Table~1. For comparison, the results of some selected theoretical calculations, such as a shell model~\cite{Kitagawa93}, microscopic cluster models~\cite{Csoto93,Baye94,Varga95,Grigorenko98}, a Hartree-Fock~\cite{Kitagawa99} and a Skyrme Hartree-Fock model~\cite{Chandel03}, in which both these nuclei were considered, are also shown. It is seen that our result for $^8$B is in agreement with the experimental value~\cite{Fan15} obtained with the modified Glauber model approach in the recent analysis of all existing data for $\sigma_{\rm R}$ but in disagreement with the earlier results of Refs.~\cite{Tanihata88} and \cite{Obuti96}. The value of the matter radius $R_{\rm m}$ deduced in the present work is also consistent with most theoretical predictions presented in Table~1. In Ref.~\cite{Grigorenko98} the theoretical description of $^8$B is performed assuming a $\left( \alpha + ^3\textrm{He} + p \right)$  three-cluster model with explicit inclusion of the binary $^7$Be + $p$ channel. The model predicts a value of $R_{\rm m}$ = 2.59~fm and correctly reproduces the experimentally observed narrow width of the momentum distribution for the $^7$Be fragments in the $^8$B high-energy breakup~\cite{Smedberg1999}. According to this model, the presence of a loosely bound proton leads to a contraction of the $^7$Be cluster inside $^8$B. This finding is confirmed in the present work. Indeed, the deduced core radius $R_{\rm c}$ = 2.25(3)~fm is smaller than the $^7$Be matter radius $R_{\rm m}$ = 2.31(2)~fm~\cite{Tanihata88}.

\begin{table}[t]
	\centering 
	\caption
	{Values of the rms point matter radii $R_\textrm{m}$ for $^8$B and $^8$Li (in fm) from experimental and theoretical studies.}
	\vspace*{2pt}
	\begin{tabular}{llll|lll}
		\hline
		\multicolumn{4}{c|}{Experiment}                          &                          \multicolumn{3}{c}{Theory}                           \\
		\hline
		\multicolumn{2}{c}{$^8$B}             &     \multicolumn{2}{c|}{$^8$Li}      & \multicolumn{1}{c}{$^8$B} & \multicolumn{1}{c}{$^8$Li} &  \\
		\hline
		2.38 (4)         & \cite{Tanihata88}  & 2.37 (2)         & \cite{Tanihata88} & 2.740                  & 2.531                    & \cite{Kitagawa93}   \\
		2.50 (4)         & \cite{Khali96}     &                  &                   & 2.57                   & 2.45                     & \cite{Csoto93}      \\
		2.43 (3)         & \cite{Obuti96}     &                  &                   & 2.73                   & 2.64                     & \cite{Baye94}       \\
		2.55 (8)         & \cite{Negoita96}   &                  &                   & 2.56                   & 2.44                     & \cite{Varga95}      \\
		2.45 (10)        & \cite{Fukuda99}    &                  &                   & 2.59                   & 2.38                     & \cite{Grigorenko98} \\
		2.61 (8)         & \cite{Fan15}       & 2.39 (5)         & \cite{Fan15}      & 2.627                  & 2.515                    & \cite{Kitagawa99}   \\
		\textbf{2.58(6)} & \textbf{this work} & \textbf{2.50(6)} & \cite{Dobrov06}   & 2.57                   & 2.54                     & \cite{Chandel03}    \\
		\hline
	\end{tabular}
\end{table}

Information on the $^8$Li nuclear matter size is rather scarce. The value of $R_{\rm m}$ obtained from the measurement of ${\rm d}\sigma/{\rm d}t$ for the elastic $p^8$Li scattering~\cite{Dobrov06} is somewhat larger than that deduced from the measured cross section $\sigma_{\rm R}$~\cite{Fan15} and within the error limits agrees with the results of most theoretical calculations displayed in Table~1. Note that the theoretical predictions~\cite{Csoto93,Varga95,Kitagawa99,Chandel03} for the sizes of both $^8$B and $^8$Li nuclei are consistent with our experimental $R_{\rm m}$ values.

A simple geometrical classification scheme was suggested~\cite{Grigorenko98} as a criterion for a quantitative assessment of halo nuclei. The ratio of the valence nucleon to the core nucleon radii $\kappa = R_{\rm v} / R_{\rm c}$ is used as a gauge for the halo existence. For light nuclei close to the valley of beta stability, theory predicts typically values of $\kappa \sim 1.20-1.25$, while for halo nuclei this value can be essentially larger, up to $\kappa > 2$~\cite{Jonson2004}. From the present results we deduce a value of $\kappa = 1.88 (14)$ for $^8$B. This value may be compared with $\kappa = 1.04 (22)$ for the nonhalo nucleus $^8$Li~\cite{Dobrov06}.
Note, that the density tails of $^8$B and $^8$Li deduced in our work and in~\cite{Fan15} have different shapes due to different model density distributions and different experimental data used in the analysis. However, our conclusion that $^8$B is a halo nucleus while $^8$Li is a nonhalo one with a noticeable neutron skin~\cite{Dobrov06} is in qualitative agreement with the findings of \cite{Fan15} and \cite{Grigorenko98} that the proton density tail in $^8$B is significantly more enhanced in comparison with the neutron tail in $^8$Li.

Under the assumption that for $^8$B the rms radius of the neutron distribution $R_{\rm n}$ is equal to the core radius $R_{\rm c}$ and using the expression
\begin{equation}
{A {R_{\rm m}}^2 = Z {R_{\rm p}}^2 + N {R_{\rm n}}^2}~,
\end{equation}
where $Z$ and $N$ are the numbers of protons and neutrons, we obtain the rms radius of the proton distribution as $R_{\rm p} = 2.76 (9)$~fm. Finally, taking into account the relation between the point proton and the charge radius of a nucleus~\cite{Lu13}, the $^8$B charge radius is deduced to be
\begin{equation}
{R_{\rm ch} ~=~ 2.89 ~\pm 0.09 ~\rm {fm}}~.
\end{equation}
The weighted average charge radii $R_{\rm ch} = 2.43 (5)$~fm and $R_{\rm ch} = 2.41 (3)$~fm for the stable $^{10}$B and $^{11}$B are known from electron and $\pi^+$ scattering measurements and from muonic atom X-rays studies~\cite{Angeli13}. The $^8$B charge radius is fairly larger than the ones for the stable boron isotopes. This observation supports the concept of a halo structure in $^8$B. The present value of $R_{\rm p}$ for $^8$B is in good agreement with some theoretical calculations: 2.75~fm~\cite{Grigorenko98}, 2.74~fm \cite{Csoto93}, and 2.73~fm \cite{Varga95,Chandel03}. For the thickness of the proton skin $\delta_{\rm pn} = R_{\rm p} - R_{\rm n}$ we deduce from the present measurements a value of 0.51(9)~fm. For $^8$Li, combining the matter radius obtained by our method~\cite{Dobrov06} with the proton radius deduced from the nuclear charge radius~\cite{Lu13}, the thickness of the neutron skin can be determined to be $\delta_{\rm np} = 0.46 (12)$~fm.

In summary, we have measured the absolute differential cross section for small-angle proton elastic scattering on the $^8$B nucleus. The measurement was performed in inverse kinematics with the secondary $^8$B beam with an energy of 0.7~GeV/u using the active hydrogen target IKAR. An analysis of the experimental data was performed on the basis of the Glauber multiple-scattering theory. A good description of the measured cross section has been obtained with four different phenomenological parameterizations of the nuclear-density distributions. The deduced rms matter radius of $^8$B is nearly identical for all parameterizations used resulting in the averaged value $R_{\rm m}$ = 2.58(6)~fm. Under the assumption that $^8$B consists of a $^7$Be core and a valence proton, the rms halo radius has been deduced to be $R_{\rm h}$ = 4.24(25)~fm, thus directly indicating a halo structure of $^8$B. This result is in close correspondence with the behaviour of the curvature in the $t$-dependence of the $\textrm{ln} \left( {\rm d}\sigma/{\rm d}t\right) $ for $^8$B. A significant positive curvature for the case of $^8$B, as compared to a negative curvature for $^8$Li, is a fingerprint for the $^8$B halo nature.  A comparison of the $R_{\rm m}$ values for both the $^8$B and $^8$Li nuclei with the results of theoretical calculations~\cite{Csoto93,Varga95,Kitagawa99,Chandel03} show a satisfactory consistency. The $^8$B proton radius $R_{\rm p}$ = 2.76(9)~fm and the corresponding charge radius $R_{\rm ch}$ = 2.89(9)~fm are determined experimentally for the first time. The deduced charge radius of the proton-rich $^8$B nucleus is significantly larger than that of nuclei of the stable isotopes $^{10}$B and $^{11}$B confirming the existence of a halo in $^8$B.

\section*{Acknowledgements}

The authors are grateful to A.~Bleile, G.~Ickert, A.Br\"{u}nle, K.-H.~Behr and W.~Niebur for their technical assistance in the preparation of the experimental set-up. The visiting group from PNPI thanks the GSI authorities for the hospitality.

\section*{References}

\bibliography{bibfile}

\end{document}